\documentclass[twocolumn,aps,floatfix,showpacs,showkeys,prb]{revtex4}

\pdfoutput=1

\usepackage{graphicx}
\usepackage{amsmath}
\usepackage{amssymb}
\usepackage{amsfonts}
\usepackage{tabularx}
\usepackage{color}

\newcommand{\ls}{$m_L/m_S$}

\begin{document}

\title{Role of symmetry in quantitative EMCD experiments}

\author{J\'{a}n Rusz} \email{jan.rusz@fysik.uu.se}
\affiliation{Department of Physics, Uppsala University, Box 530, S-751~21 Uppsala, Sweden}
\affiliation{Institute of Physics, Academy of Sciences of the Czech Republic, Na Slovance 2, CZ-182 21 Prague, Czech Republic}




\date{\today}

\begin{abstract}
We have analysed a number of experimental geometries (two and three-beam geometry, general systematic row setup, zone axis orientation) for the electron magnetic circular dichroism measurements (EMCD). The effect of asymmetries in those setups is described in detail, explaining the advantages of the three-beam geometry. We observe a complicated thickness and orientation dependence of the dichroic signal and asymmetries. Therefore we stress the importance of the simulations, which should accompany the EMCD experiments.
\end{abstract}


\maketitle

\section{Introduction}


The recent discovery of the possibility to observe the magnetic circular dichroism in the transmission electron microscope\cite{nature} opened a new route to atom-specific magnetic studies with an unprecedented spatial resolution. This technique was named electron magnetic circular dichroism (EMCD). Although principle of this technique shares many common points with x-ray magnetic circular dichroism (XMCD), dynamical electron diffraction has a pronounced effect on the observed spectra. This brings many subtleties for the interpretation of experiments and the extraction of physical quantities. In order to turn the technique into a quantitative probe of magnetic properties, it is thus necessary to analyse possible sources of inaccuracies connected with dynamical diffraction. We will deal with so far the only viable EMCD approach, in which the crystal itself is used as beamsplitter\cite{Micha} and the EMCD signal is obtained as a difference of spectra measured at two detector positions. We concentrate here on the role of symmetry of the experimental setup geometry, i.e., relative orientation of incoming beam vs sample orientation vs detector positions. Note that yet before the EMCD was discovered, signals of odd and even symmetries were separated in an electron energy loss experiment\cite{batson}, though relation between the signal of odd symmetry and magnetism were not yet noticed.


In a recent publication\cite{prbtheory} we have described simulations of electron energy-loss near edge structures (ELNES). This theory is based on the Bloch wave approach\cite{metherell,nelhiebel,KohlRose85,peng} to the dynamical diffraction effects, including higher-order Laue zones. Matrix elements of inelastic transitions are evaluated from first-principles electronic structure. This method was applied to study EMCD in the experimental geometry proposed in the original paper\cite{nature}. Those simulations were done within the systematic row approximation. Inclusion of high order Laue zones provides more complete, richer description leading to qualitatively new observations presented in this work.

An important theoretical development for quantitative use of the EMCD technique was the derivation of sum rules\cite{oursr,lionelsr}, which relate the measured spectra to spin ($m_S$) and orbital ($m_L$) magnetic moments of a studied atom in the sample. Among the first quantitative measurements of the \ls{} ratio are Refs.~\onlinecite{lionelsr,schatt} and, particularly, Ref.~\onlinecite{klaus}, where a method of data treatment and statistical analysis was described.

The sum rules require two measurements in symmetrical conditions for determination of the orbital to spin moments ratio. When the symmetry condition is fulfilled, sum rules revealed a flexibility of the EMCD technique allowing to measure the dichroic signal in the whole diffraction plane. Consequences of these findings are the main points of this paper.

After describing computational methods (Sec.~\ref{sec:calc}) we study in detail the originally proposed experimental geometry (Sec.~\ref{sec:2bc}) - the two-beam case (2BC). Maps of the dichroic signal are presented. We demonstrate that 2BC actually does not provide two strictly symmetric measurements and analyse the error introduced to the extracted \ls{} ratios. An improved experimental setup based on three-beam case (3BC) geometry is discussed in Sec.~\ref{sec:3bc}. In Sec.~\ref{sec:sys} we study the general systematic row geometry. Finally, in Sec.~\ref{sec:za} we analyze the zone axis (ZA) orientation and its suitability for EMCD measurements.

\section{Method of calculations\label{sec:calc}}

In Ref.~\onlinecite{prbtheory} we have described our theoretical approach to simulate general orientation-sensitive ELNES experiment from first principles. These methods will be fully utilized in this work. Here we present calculations for bulk bcc Fe including higher order Laue zones, Weickenmeier-Kohl scattering potentials\cite{WeickKohl} and Debye-Waller factor 0.35\AA$^2$ for Fe at room temperature\cite{dwfe}. The energy of the electron beam was set to 300keV and the sample thickness varied between 5~nm and 50~nm. For solution of the eigenproblem for both incoming and outgoing beam we use basis consisting of about 700 beams (maximum $w_\mathbf{g} \equiv \xi_\mathbf{g} s_\mathbf{g} = 10^5$, where $\xi_\mathbf{g}$ and $s_\mathbf{g}$ are the extinction distance and excitation error, respectively).

Compared to our previous work, Ref.~\onlinecite{prbtheory}, we have significantly improved the summation over the Bloch coefficients. Details of implementation and study of the convergence are given in Appendix \ref{sec:dd12}, here we list only the convergence criteria. A dipole-type MDFF asymptotics is assumed in estimating the term magnitudes and all terms exceeding $10^{-5}$ are included in the summation. For zone axis simulations an increased limit $3 \times 10^{-5}$ was used to reduce computational costs.


For calculation of matrix elements $S(\mathbf{q},\mathbf{q'},E)$ - the mixed dynamic form factors (MDFF)\cite{KohlRose85} - we used density functional theory. Spin-orbit interaction was included and magnetic moments were aligned with incoming beam direction. Particularly, $\mathbf{M}||(016)$ to simulate $\sim\!\!10^\circ$ tilt of the sample within systematic row setups with $\mathbf{G}=(200)$. In zone axis simulations, $\mathbf{M}||(001)$. We use Rayleigh expansion of the Coulomb operator to the first order ($\lambda=\lambda'=1$ approximation, where $\lambda$ is the order of the terms included in expansion), which keeps the computational efficiency of dipole approximation while producing much more accurate results for larger momentum transfer vectors.
We refer the reader to the above-mentioned paper\cite{prbtheory} for more details on the simulations.


\section{Two-beam case\label{sec:2bc}}


\begin{figure}[t]
  \includegraphics[width=5.5cm]{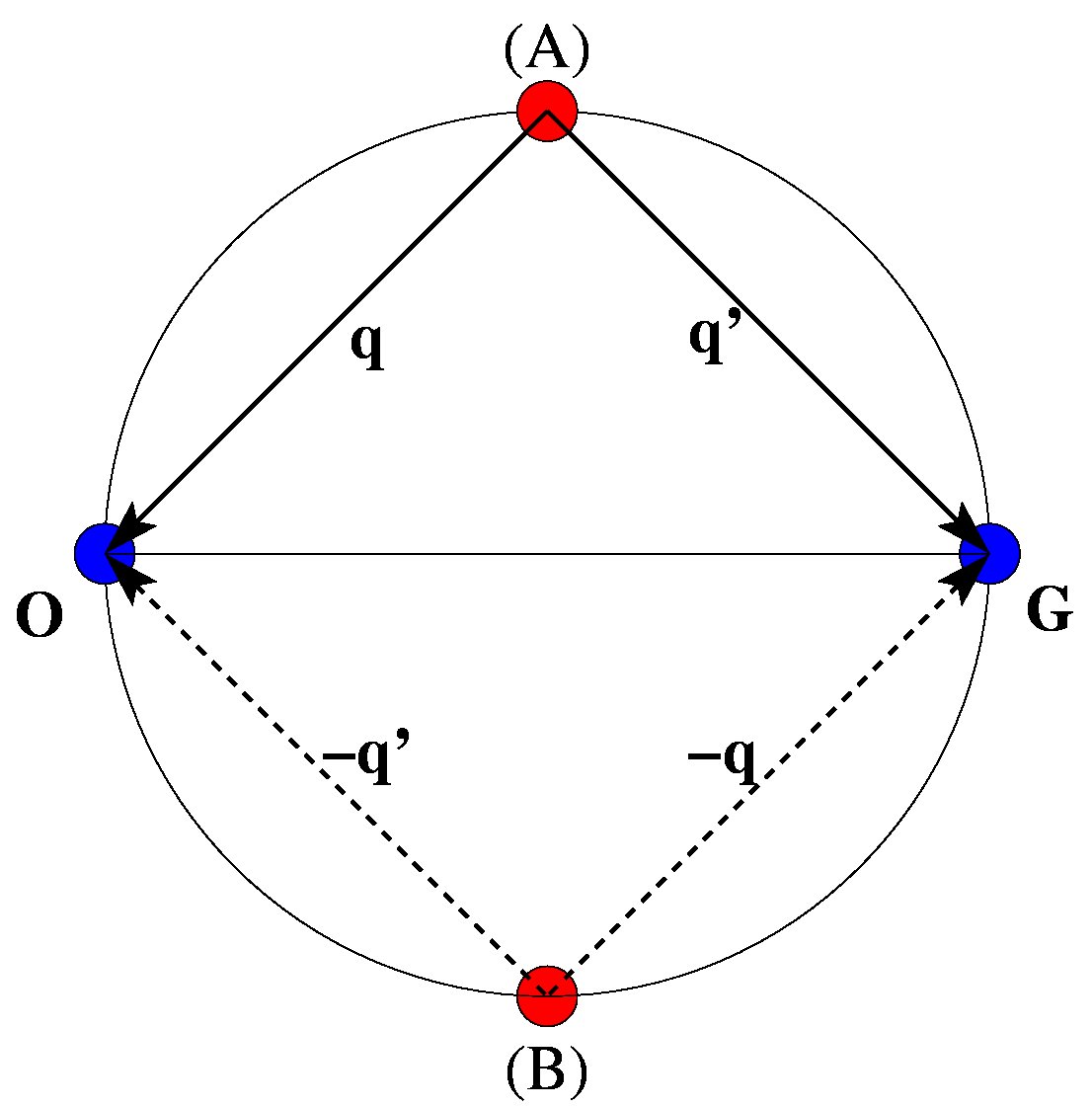}
  \caption{(color online) The Thales circle circumscribed to excitations $\mathbf{0}$ and $\mathbf{G}$ is a set of detector positions, for which the two momentum transfer vectors $\mathbf{q}$ and $\mathbf{q'}$ are perpendicular. The detector positions denoted as (A) and (B) fulfill also $|\mathbf{q}|=|\mathbf{q'}|$. These two detector orientations are usually quoted as \emph{the Thales circle positions}.
  \label{fig:thales}}
\end{figure}

\begin{figure}[t]
  \includegraphics[width=6cm]{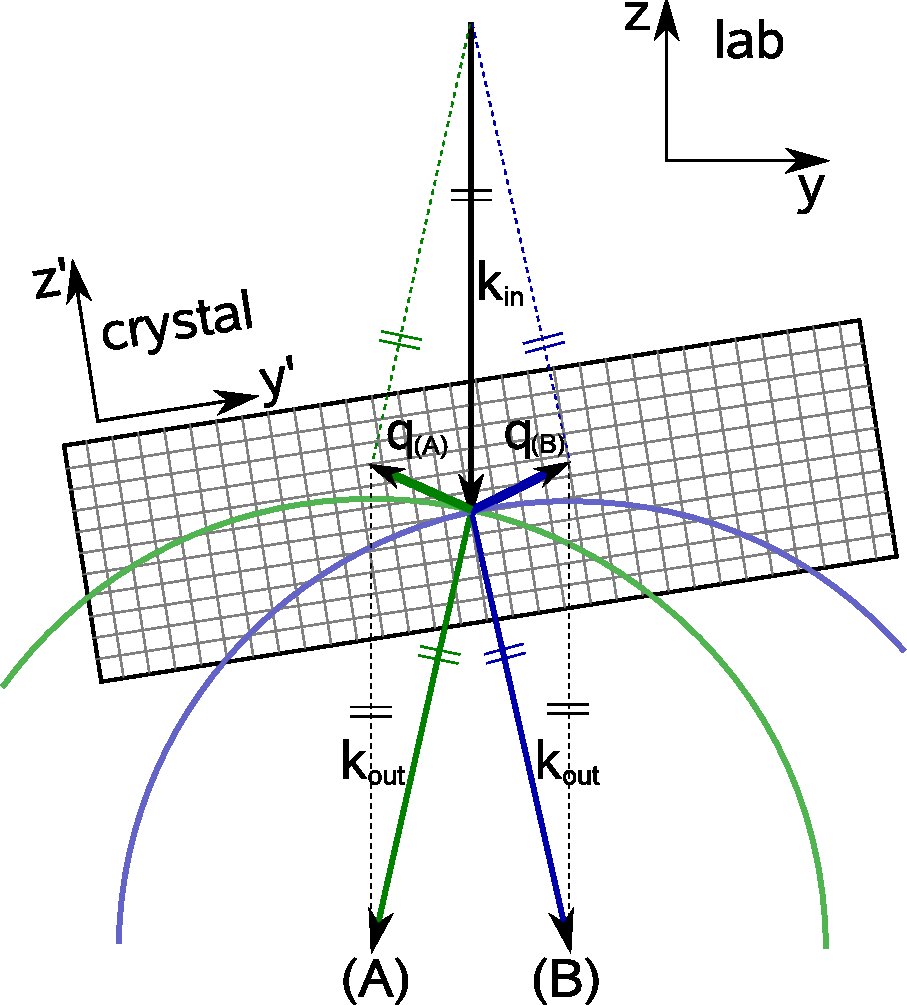}
  \caption{(color online) Asymetry of the two-beam case. The incoming beam with wavevector $\mathbf{k}_\mathrm{in}$ is tilted with respect to the zone axis, which is perpendicular to the sample surfaces. The square grid is a sketch of the crystal lattice. Detected outgoing beams for two detector positions (A) and (B) are described by their respective wavevectors $\mathbf{k}_\mathrm{out}$. The dotted light arrows are auxiliary vectors parallel with incoming and outgoing beam wavevectors used for construction of momentum transfer vectors $\mathbf{q}_\mathrm{(A)}$ and $\mathbf{q}_\mathrm{(B)}$ (thick arrows inside the crystalline lattice).\label{fig:assym2bc}}
\end{figure}

We will describe in more detail the originally proposed geometry of the EMCD experiment, as applied in Refs.~\onlinecite{nature,prbtheory}. This geometry is based on exciting a systematic row of reflections (determined by the reciprocal lattice vector $\mathbf{G}$) by tilting the incoming beam $\mathbf{k}_{in}$ a few degrees from some highly symmetric direction, such as $(001)$, within a plane perpendicular to the selected $\mathbf{G}$. Then the incoming beam is tilted a few miliradians further along the systematic row in order to produce a two beam case (see Fig.~3 of Ref.~\onlinecite{prbtheory}). This is achieved by setting the Laue circle center of the incoming beam to $lcc=\mathbf{G}/{2}$. We define momentum transfer vectors $\mathbf{q}=\mathbf{k}_{out}-\mathbf{k}_{in}$ and $\mathbf{q'}=\mathbf{q}+\mathbf{G}$, with respect to the transmitted and Bragg scattered beams, respectively. By taking the excitations $\mathbf{0}$ and $\mathbf{G}$ as the diameter of a Thales circle, then for all detector positions $\mathbf{k}_{out}$ lying on this circle, the projections of $\mathbf{q}$ and $\mathbf{q'}$ on the diffraction plane are perpendicular. A particularly symmetric situation is achieved, when $|\mathbf{q}|=|\mathbf{q}'|$, i.e., the detector placement in a direction corresponding to the top (A) or bottom (B) point of the Thales circle, see Fig.~\ref{fig:thales}. In a simple two-beam approximation, measuring at these detector positions is formally in the closest analogy to XMCD experiments.\cite{nature} The archetype EMCD experiment thus amounts to acquiring two spectra at these two detector positions. The difference of these spectra is attributed to the magnetic circular dichroism, because in the presence of inversion symmetry the MDFF changes sign of its imaginary part, $S(\tilde{\mathbf{q}},\tilde{\mathbf{q}}',E) \equiv S(-\mathbf{q'},-\mathbf{q},E)=S^\star(\mathbf{q},\mathbf{q'},E)$,\cite{nature} when moving from detector position (A) to (B), see Fig.~\ref{fig:thales}. The real part is assumed to stay unchanged and cancel out in the difference. Then the difference spectra would depend only on \textit{imaginary} parts of contributing MDFFs. Energy integral of the imaginary part of MDFF was shown to be proportional to magnetic moments of the studied atom\cite{oursr}.

Now we will demonstrate, that actually the (A) and (B) detector positions are, strictly speaking, not symmetric and therefore the signal measured at these two positions differs more than solely by reversing the sign of imaginary parts of contributing MDFFs. 

The assymetry itself appears due to the tilt of the crystal (usually approx.\ 5 to 10 degrees). The situation is schematically drawn in Fig.~\ref{fig:assym2bc}. Note that 1) the outgoing beam wave vectors $\mathbf{k}_\mathrm{out}$ for the two detector positions have different relative orienation with respect to the crystal lattice; 2) the same is valid for the momentum transfer vectors $\mathbf{q}_{(A)},\mathbf{q}_{(B)}$ drawn inside the sample. These two facts are reflected in two sources of asymmetries: 1) different reciprocal wave Bloch fields at detector position (A) vs detector position (B), and, 2) difference in calculated MDFFs, which goes beyond the mere change of the sign of the imaginary part.

The difference in outgoing Bloch wave fields appears because of the different relative orientations of the Ewald spheres for observed outgoing beams with respect to the reciprocal lattice. Note though, that in the systematic row approximation (SRA), i.e., when we take into account only Bragg reflections along the $x$-axis (perpendicular to the Fig.~\ref{fig:assym2bc} plane), then the two eigensystems become equal. In such case, from the whole reciprocal lattice we consider only the systematic row and such simplified geometry acquires a mirror symmetry plane defined by $x$ and $\mathbf{k}_{in}$. Nonetheless, this is an approximation and it is necessary to analyse its validity and appropriateness.\footnote{Since crystal is tilted the outgoing beam directions at detector positions (A) and (B) have different components perpendicular to the sample surface and thus the elongations $\gamma^{(j)}$ of Bloch-wave vectors $\mathbf{k}^{(j)}$ will differ slightly, see Eq.~(7) in Ref.~\onlinecite{prbtheory}. This causes a weak asymmetry in scattering cross-sections even within systematic row approximation. Its influence depends on sample thickness and can be seen in 2BC maps of the \ls{} ratio, Fig.~\ref{fig:ls23}.}

\begin{figure}[t]
  \includegraphics[width=8.5cm]{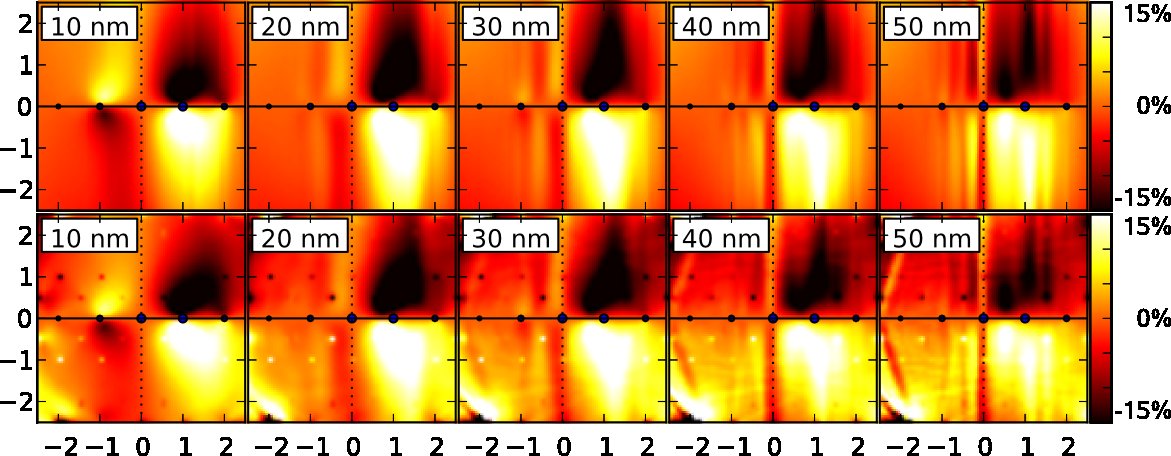}
  \caption{(color online) Up-down relative difference maps of Fe-$L_3$ edge calculated in 2BC as a function of thickness within (top row) systematic row approximation, and (bottom row) including HOLZ. The $x$-axis coincides with the systematic row and the origin of the coordinate system is at the position of the transmitted beam (in the middle). Tick labels of axes are multiples of the Bragg-reflection $G_{(200)}$. The same area of the diffraction plane is used in all subsequent figures. \label{fig:2bcsraholz}}
\end{figure}

Strength of the latter effect depends on the anisotropy of the electron distribution surrounding the excited atoms. Note that this error also increases with larger energy losses, since larger energy loss shortens the $\mathbf{k}_\mathrm{out}$ more and the angle between related momentum transfer vectors increases. The effect itself introduces differences between the real parts of MDFFs, which were expected to be symmetry-related. As a consequence, the difference signal from positions (A) and (B) will contain not only the imaginary parts (the magnetic signal) but also the real parts of MDFFs.


Figure~\ref{fig:2bcsraholz} compares ``up-down'' difference maps calculated within the SRA with calculation including higher order Laue zones (HOLZ calculation), Sec.~\ref{sec:calc}. They display the \emph{antisymmetric part} of the signal with respect to the systematic row axis - i.e., the difference of the diffraction pattern and its mirror image with respect to the systematic row. We will call them simply \emph{difference maps}, contrary to names `dichroic maps' or `dichroic signal' used in recent literature. The reason is that in the majority of the experimental geometries this difference signal is only an approximation to the magnetic component. Due to asymmetries these maps contain also a fraction of non-magnetic signal, which can introduce a systematic error into the \ls{} evaluation. \emph{Relative difference maps} are difference maps nomalized by the symmetric part of the signal, i.e., the difference divided by the sum.

There are several local maxima and minima in the diffraction plane, Fig.~\ref{fig:2bcsraholz}. Their positions and strengths depend on thickness and, in general, none of them is on the Thales circle. Remembering that the sum rules allow \ls{} extraction at any place in diffraction plane, it is obvious that Thales circle positions are not the optimal detector position for measurement, when it comes to optimizing signal-to-noise ratio (SNR). Actually, there is no general `optimal detector position'. The optimal detector position is material, thickness and orientation dependent (see also Appendix~\ref{sec:dichmaps}). This problem was studied in detail in Ref.~\onlinecite{snr} from the SNR point of view.

The main features of HOLZ calculation are reproduced by the SRA, see Fig.~\ref{fig:2bcsraholz}. However, there are some qualitative changes in the HOLZ simulation, which are not present within SRA and the deviations increase with thickness. Two pronounced features appear in the HOLZ maps, which are not present in the SRA maps: Kikuchi lines and weakly excited Bragg spots away from the systematic row. A Kikuchi line and its mirror image can be best seen at thicknesses 40~nm and 50~nm -- the skew lines in the upper left and lower left corners of HOLZ maps. The weakly excited Bragg spots are almost invisible in diffraction patterns, but the relative difference maps enhance their visibility. Presence of both these effects in relative maps is a sign of the asymmetry.



\begin{figure}[t]
  \includegraphics[width=8.5cm]{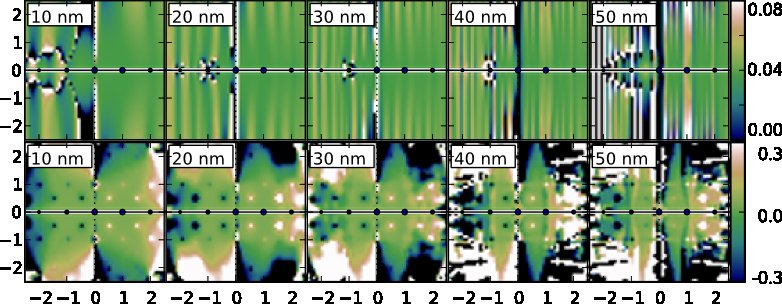}
  \caption{Apparent \ls{} ratio maps in the 2BC geometry as a function of thickness calculated within two models: systematic row approximation (SRA, top row) and full calculation (bottom row).\label{fig:ls23}}
\end{figure}

If these maps would originate solely from imaginary parts of MDFFs (a pure magnetic signal), it should be possible to obtain \ls{} ratio at any position in the diffraction plane. This follows from the form of spin and orbital sum rules expressions, according to which the dynamical diffraction effects cancel out in the \ls{} ratio. We tested this by evaluating the sum-rules expression\cite{oursr,lionelsr}
\begin{equation} \label{eq:ls}
  \frac{m_L}{m_S} = \frac{2}{3} \frac{\Delta_{L_3}+\Delta_{L_2}}{\Delta_{L_3}-2\Delta_{L_2}}
\end{equation}
throughout the diffraction plane (here $\Delta_{L_{2,3}}$ is the difference signal integrated over $L_2$ or $L_3$ edge, respectively). On top of the usual sum rules approximations,\cite{ebert} by writing this expression we assume that i) both spin and orbital momentum vectors are parallel and ii) the magnetic dipole term is negligible compared to the spin moment.

The results of these simulations are in the Fig.~\ref{fig:ls23}, where we compare such evaluation within SRA and HOLZ calculation. The ``\ls{} ratio'' is by no means constant in the diffraction plane, not even in SRA. Therefore, in the following text, we will refer to values in these maps as \textit{apparent \ls{} ratio} to distinguish them from the \ls{} ratio obtained by electronic structure calculations, which is 0.045 for the energy integration range used in our simulations ($1.5-2.0$eV above the Fermi level, see Appendix \ref{sec:eint}). 

There are several factors that affect the \ls{} ratio: 1) the assumption of equal radial wavefunctions for both $2p_{1/2}$ and $2p_{3/2}$ initial states made in derivation of sum rules, 2) slight difference in the direction of the spin and orbital moments due to magneto-crystalline anisotropy, 3) magnetism of $s$-states and related $p \to s$ transitions and $p \to s/p \to d$ cross-terms in inelastic transition matrix elements, and, 4) the asymmetry discussed above.

The first effect can be easily quantified by performing a test calculation with equal initial-state radial wavefunctions for both edges. The fact that $p_{1/2}$ and $p_{3/2}$ radial wavefunctions are slightly different was observed to decrease the apparent \ls{} ratio by $\sim 0.01$ in the areas weakly affected by other sources of deviations. This explains the mean value of 0.035 within regions close to the systematic row. 

The second effect depends on the strength of magneto-crystalline anisotropy, i.e., to what extent the $\langle\hat{\mathbf{L}}\rangle$ and $\langle\hat{\mathbf{S}}\rangle$ follow the external magnetic field and, particularly, what is the angle between $\langle\hat{\mathbf{L}}\rangle$ and $\langle\hat{\mathbf{S}}\rangle$. Considering the vector nature of sum rules\cite{oursr}, prefactor of the spin and orbital magnetic moments is actually a vector ($\mathbf{p}$) and thus the ratio of type $\mathbf{p}\cdot\langle\hat{\mathbf{L}}\rangle/\mathbf{p}\cdot\langle\hat{\mathbf{S}}\rangle$ may vary with varying $\mathbf{p}$ if $\langle\hat{\mathbf{L}}\rangle \nparallel \langle\hat{\mathbf{S}}\rangle$. This is a tiny effect in bcc-Fe, which has extremely low magnetocrystalline anisotropy. As a result, at $\mathbf{M} \parallel (016)$ the angle between spin and orbital moments is below $0.1^\circ$ and therefore the error thus introduced is of no practical importance. For systems with strong magnetocrystalline anisotropy this effect might become visible. That might, in principle, allow to determine the angle between spin and orbital magnetic moment by following the changes of the \ls{} ratio throughout the diffraction plane. However, detailed study of this effect goes beyond the scope of the present paper.

The asymmetry effect is the most crucial one. It can lead to a quite large systematic error, when the \ls{} value is extracted directly from 2BC maps further from the systematic row.


The asymmetry within the SRA has a form of quasi-periodic pattern of vertical stripes. These modulations have a long period and low amplitude at low sample thicknesses, but as the thickness increases, the period shortens and amplitude increases. This effect can be traced doen to differences of elongations $\gamma^{(j)}$ of outgoing beam Bloch waves in upper and lower half-planes, which is caused by difference of projections of the outgoing wave vectors $\mathbf{k}_{(A)}$ and $\mathbf{k}_{(B)}$ to the sample surface normal (see also footnote in discussion of asymmetry under Fig.~\ref{fig:assym2bc}). Thickness enters the differential scattering cross-section through factors\cite{prbtheory} $e^{i(\gamma^{(j)}-\gamma^{(l)})t}$, which suggests that at larger thickness the scattering cross-section will be more sensitive to differences in elongations $\gamma^{(l)}$.

The asymmetry within the full HOLZ calculation is stronger and more complex. Notice, for example, that at the position of Kikuchi lines the apparent \ls{} ratio is estimated to be off by an order of magnitude. Deviations are stronger at detector positions, where the magnetic signal becomes weaker. Also, the deviations increase with increasing thickness, which puts an upper limit on sample thicknesses for EMCD measurements. To summarize, it is the asymmetry originating in the complete treatment of the dynamical diffraction, which has the strongest influence on observed \ls{} ratios in the 2BC. A more detailed computational study of the asymmetry in the 2BC can be found in Ref.~\onlinecite{2bcasymm}.


Still, there is a quite large area in the diffraction plane, where the 2BC geometry allows accurate \ls{} ratio measurement, Fig.~\ref{fig:ls23}. Though, due to dynamical diffraction effects, the shape and position of this area is thickness dependent. Therefore it would be of advantage to accompany experiments in 2BC setup with simulations. Eventually, one can perform experiments at varying sample thicknesses to identify regions with biased \ls{} values. If a suitable thickness and detector position in diffraction plane is identified, this geometry might be preferred for spectroscopic measurements of EMCD, because the two-beam orientation concentrates the magnetic signal to the area inside and close to the Thales circle (compare with three-beam orientation, Sec.~\ref{sec:3bc}).



\section{Three-beam case\label{sec:3bc}}

\begin{figure}[t]
  \includegraphics[width=8cm]{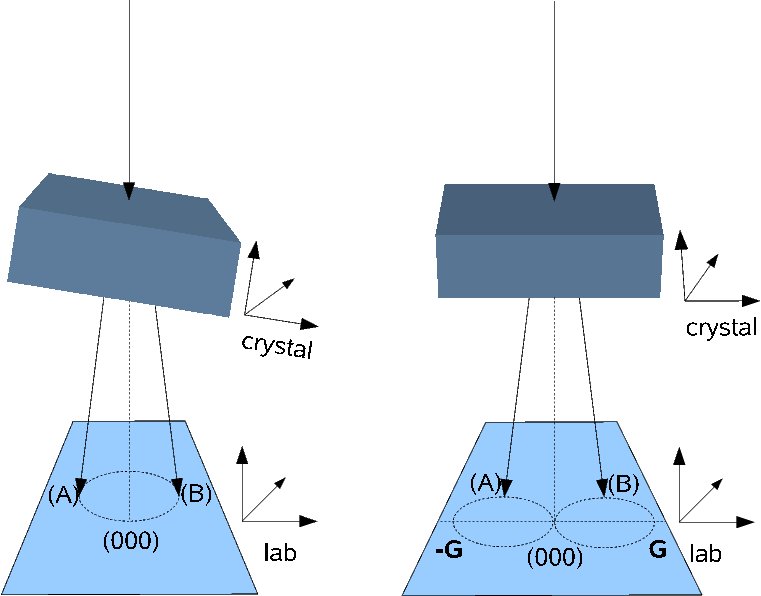}
  \caption{(color online) Default 2BC (left) and proposed 3BC (right) experimental geometries showing a tilted sample
           and detector positions (A) and (B).\label{fig:assym}}
\end{figure}

\begin{table}[bth]
  \begin{tabular}{lccc}
                        &       default       &              alt.\ 1           &       alt.\ 2                  \\
    \hline \hline
    \vspace{2mm}
    lcc$_{in}$          & $(\frac{G}{2} 0 0)$ &             $(000)$            &      $(000)$                   \\
    \vspace{2mm}
    det$_{out}^{(A)}$   & $(\frac{G}{2} \frac{G}{2} 0)$     & $( \frac{G}{2} \frac{G}{2} 0)$    & $(\frac{G}{2}\bar{\frac{G}{2}} 0)$ \\
    \vspace{2mm}
    det$_{out}^{(B)}$   & $(\frac{G}{2}\bar{\frac{G}{2}}0)$ & $(\bar{\frac{G}{2}}\frac{G}{2}0)$ & $(\bar{\frac{G}{2}}\bar{\frac{G}{2}}0)$ \\
    \hline
  \end{tabular}
  \caption{Parameters for 2BC and 3BC geometries and Thales circle detector positions. All are given in terms of 
           systematic row vector $\mathbf{G}$ and Laue circle center ($lcc$) for incoming beam and detector position 
           assuming the $x$ axis parallel to $\mathbf{G}$ and $z$ axis parallel to the zone axis. In 3BC there are two 
           possibilities for outgoing beams - `above' (alt.\ 1) or `below' (alt.\ 2) the systematic row, respectively. 
           The right part of Fig.~\ref{fig:assym} shows alt.\ 1.\label{tab:lcc}}
\end{table}

\begin{figure}[t]
  \includegraphics[width=8.6cm]{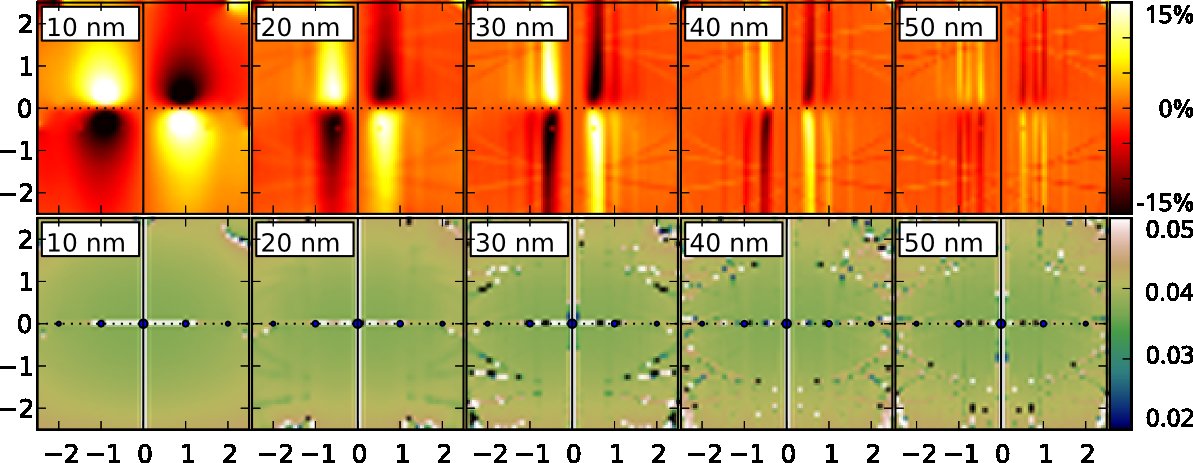}
  \caption{(color online) Top row: maps of the $L_3$ relative difference signal of bcc-Fe in the 3BC orientation constructed using the vertical mirror axis (solid line). Systematic row is shown as a dashed line and Bragg spots as small (blue) circles. Bottom row: the \ls{} ratio map constructed using the vertical mirror axis.\label{fig:3bc}}
\end{figure}

\begin{figure}[t]
  \includegraphics[width=8.6cm]{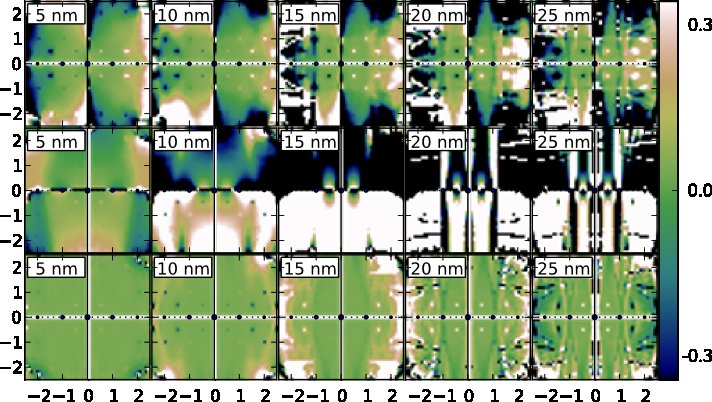}
  \caption{Map of the \ls{} ratio in the slightly misoriented 3BC, $lcc=(0.05,0,0)=0.025G_{(200)}$. Top row: up-down difference (horizontal mirror axis), middle row: left-right difference (vertical mirror axis), bottom row: double difference (both mirror axes). Values outside the range $(-0.5,0.5)$ are white (if above 0.5) or black (if below -0.5), respectively.\label{fig:lsdd}}
\end{figure}

As we have shown, the two-beam case does not allow two independent strictly symmetrical measurements as required by the sum rules. An alternative experimental geometry, the three-beam case (3BC), is in principle free of these inaccuracies. Instead of exciting a 2BC using the incoming beam we set the Laue circle center of the incoming beam to $(000)$. This geometry is conventionally called a {\it three-beam case}, because in the diffraction plane one sees a strong spot due to the transmitted beam and two $\mathbf{G}$ and $-\mathbf{G}$ Bragg spots of equal intensity. Other spots are typically much weaker. The 3BC geometry was applied in recent experiments.\cite{klaus,snr,emcd2nm}

If there exists a mirror symmetry plane \emph{normal} to the systematic row (cf. mirror plane \emph{containing} systematic row in 2BC), this setup allows for two completely symmetric measurements. Because the mirror symmetry operation $\hat{\mathrm{M}}$ changes sign of pseudovectors, $\hat{\mathrm{M}}\mathbf{q} \times \hat{\mathrm{M}}\mathbf{q'} = - \hat{\mathrm{M}}(\mathbf{q}\times\mathbf{q'})$, the imaginary parts of MDFFs will change sign when going from (A) detector position to (B), see Fig.~\ref{fig:assym}. The eigenvalue problems for the two detector positions are equivalent, because now the incoming beam remains within the mirror plane and the outgoing beams together with their associated Ewald spheres are mutually symmetric with respect to this mirror plane.

Although the two-beam approximation and Thales circle construction for the 2BC is an oversimplified model, it was very useful for getting an insight into the EMCD phenomenon and to develop formal analogies with XMCD technique\cite{nature}. It is thus instructive to show that a similar model can be constructed also for the 3BC. In analogy with two-beam approximation for 2BC geometry, here we approximate the incoming beam by single plane wave (one-beam approximation) and set detector positions so that the outgoing Bloch field can be described by the two-beam approximation. It is possible to introduce similar Thales circle construction like in 2BC geometry, see fig.~\ref{fig:assym}, and follow the analogy with XMCD in the same way as in the 2BC. The only difference is, that the `mirror axis' is vertical and there are two Thales circles circumscribed to $\mathbf{0}$ and $\mathbf{G}$ or $\mathbf{0}$ and $-\mathbf{G}$, respectively. Laue circle centers and Thales circle detector positions for both geometries are summarized in Table~\ref{tab:lcc}. 

As mentioned above, the mirror symmetry plane cuts the diffraction plane in a direction perpendicular to the systematic row direction. Therefore to utilize the symmetry we evaluate difference maps of \emph{left and right} diffraction half-planes (cf. upper minus lower half-plane in 2BC). If we would have perfect systematic 3BC orientation, under dipole (or $\lambda=1$) approximation this difference map would contain solely a magnetic signal. This is apparent from Fig.~\ref{fig:3bc}. The \ls{} maps are very smooth with only small variations of the apparent \ls{} ratio throughout the plane. These variations are caused predominantly by $p \to d/p \to s$ transition cross-terms\cite{ebertprb,wu}. Larger variations may appear at some particular thicknesses and positions - these are caused by reduced dichroic signal due to dynamical effects (typically these are borderlines between areas of positive and negative dichroic signal). At these positions the \ls{} ratio is evaluated as a ratio of two very small numbers, i.e., becomes sensitive even to tiny deviations.

In practical measurements, though, one never reaches a perfect 3BC orientation. Either $\mathbf{G}$ or $-\mathbf{G}$ is slightly more excited than the other. Therefore it is instructive to test, how sensitive is this setup to small deviations. A combined theory-experiment study of this effect can be found in Ref.~\onlinecite{lsfollow}. Here we summarize our findings and present additional simulation results (see also Appendix~\ref{sec:dd}).

By setting $lcc_{in}$ to $(0.05,0,0)$ (approximate tilt by $0.017^\circ=1'$) we introduce an asymmetry between the left and right diffraction half-plane. Thus the `left-right' difference will not be a purely magnetic signal anymore and therefore evaluated \ls{} ratio will contain a systematic error. This error is surprisingly large, it even exceeds the asymmetry in 2BC, see Fig.~\ref{fig:lsdd} (middle row). One of its consequences is that inside the Thales circle the apparent \ls{} ratio in the upper half-plane is different than in the lower half-plane.

Evaluation of the `up-down' difference leads to a result comparable with 2BC: there are areas of the diffraction plane, where the \ls{} ratio is accurately reproduced by the apparent \ls{} ratio, Fig.~\ref{fig:lsdd} (top row).

A quite natural idea is to perform a \emph{double difference} procedure, i.e., to construct a `left-right' difference map from the `up-down' difference map. This procedure was for the first time applied in Ref.~\onlinecite{klaus}. In the slightly tilted 3BC the correction works very well in a large part of diffraction plane, Fig.~\ref{fig:lsdd} (bottom row). Note that the color variations are significantly reduced, meaning that the \ls{} ratio can be extracted from larger part of the diffraction plane. Double difference procedure takes an advantage of the observation, that the change of the signal intensity caused by the tilt, is almost the same in the upper and lower half-plane. More detailed analysis of the error correction in double difference maps is given in Appendix \ref{sec:dd}.

Overall, the 3BC orientation provides clean magnetic signal and might become the prefered geometry for quantitative EMCD measurements, when energy resolved maps of the dichroic signal are available. The small deviations from a perfect 3BC orientation, can be corrected by performing a double difference map, thus utilizing information from all four quadrants.

\section{General systematic row conditions\label{sec:sys}}

\begin{figure}[t]
  \includegraphics[width=8.6cm]{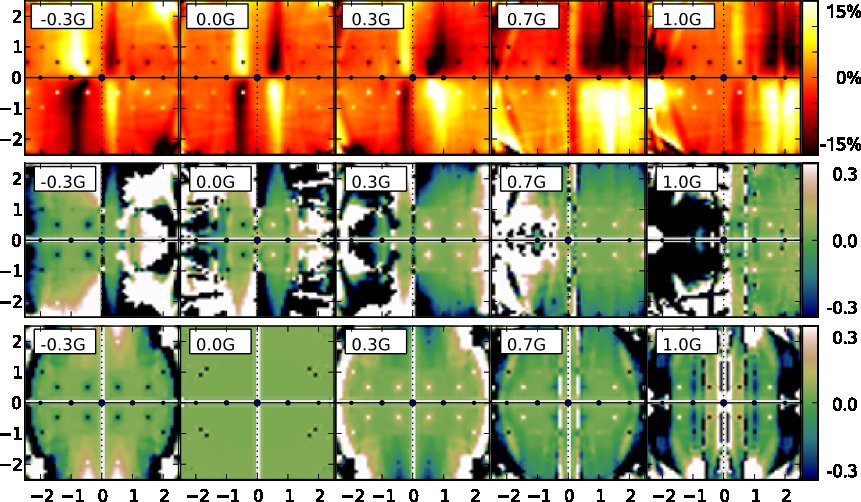}
  \caption{Maps of the `up-down' relative difference at $L_3$ (top row) of 20nm thick sample of bcc-Fe as a function of the Laue circle center (indicated in the upper left corner of each panel). The corresponding \ls{} ratio is in the middle row. A double difference map of the \ls{} ratio is shown in the bottom row.\label{fig:gensr}}
\end{figure}

A natural generalization of the 2BC ($lcc=\mathbf{G}/2$) and 3BC ($lcc=\mathbf{0}$) systematic row geometries, where one studies `up-down' difference maps, would be a general systematic row setup with in principle arbitrary $lcc=\eta\mathbf{G}$. We performed a set of calculations with $\eta$ ranging from $-1/2$ (`negative' 2BC) to $1$ with step $0.1$. 

Qualitative behavior of negative $\eta$ is similar to related positive $\eta$ values. The mirror symmetry between the geometrical setups for $\pm\eta$ lead to the following relations between resulting diffraction patterns: the non-magnetic part of the signal, which originates from real parts of MDFFs, does not change under mirror symmetry operation, but the magnetic part of the signal changes (due to its pseudo-vector nature). Therefore, in theory, if we would have a separation of the diffraction pattern into its non-magnetic part (Re[MDFF]) and magnetic part (Im[MDFF]) at certain $\eta$, we can obtain the diffraction pattern for $-\eta$ by mirroring both components, changing the sign of the magnetic component and adding them together. As a result, however, the total signals for $\pm\eta$ are not mirror images of each other and the difference is caused by magnetism. That actually provides another way, how the magnetic signal could be measured: first we measure a spectrum at tilt $\eta\mathbf{G}$ and detector position $(k_x,k_y)$ and second spectrum at tilt $-\eta\mathbf{G}$ and detector position $(-k_x,k_y)$. Their difference would be a pure EMCD signal. This scheme could be also extended to a double-difference-like procedure by measuring several spectra at $(\pm k_x, \pm k_y)$ detector positions for both tilts.

In Fig.~\ref{fig:gensr} we show difference maps of $\eta=-0.3, 0, 0.3,0.7$ and $1$. Also the apparent \ls{} maps are included. From these maps one can deduce that the `exact 2BC geometry' does not play any special role, contrary to 3BC where an additional mirror plane can be used. Indeed, for `up-down' difference maps the value of $lcc$ is--within certain limits--not important. 
By tilting the beam, the area with low influence of asymmetry moves along with the tilt. However, the thickness dependence does not allow any general conclusions about optimal detector positions (see also Appendix~\ref{sec:dichmaps}).

In the bottom row of the Fig.~\ref{fig:gensr} we show the \ls{} ratio map evaluated using the double difference procedure. A somewhat surprising conclusion can be made: the double difference procedure seems to work reasonably well also for considerable tilts. Even at $\eta=0.7$ the double difference procedure enlarges the area with low influence of asymmetry. Note also the similarity between $\eta = \pm 0.3$. Maxima at $\eta=0.3$ coincide with minima at $\eta=-0.3$ and vice-versa. The reason is that the asymmetric component of the signal changes sign, as can be expected from the symmetry considerations about the non-magnetic part discussed above. If the double difference procedure would be applied on a sum of diffraction patterns at $\eta=-0.3$ and $\eta=0.3$, the asymmetric component would cancel out.

\section{Zone axis conditions\label{sec:za}}

\begin{figure}[t]
  \includegraphics[width=8.5cm]{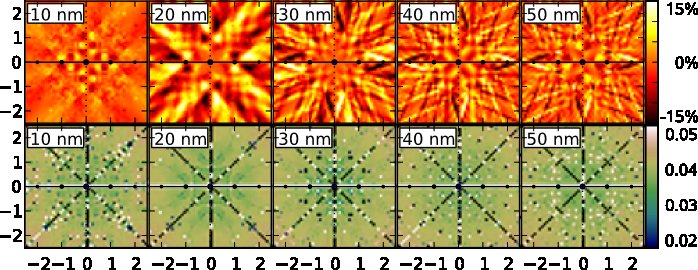}
  \caption{(color online) The top row shows maps of the $L_3$ relative difference signal of bcc-Fe in the zone axis orientation as a function of thickness. Bottom row shows corresponding \ls{} maps.\label{fig:za3bc}}
\end{figure}

Up to now, all proposed experimental setups were devised in a way to reduce the dynamical effects as much as possible--in order to make the effect physically more transparent. Intuitively, one might expect that a large number of contributing MDFFs leads to concurrent phase shifts and thus to the reduction of the net magnetic signal. Tilting the sample reduces the number of Bragg spots in the diffraction plane and thus reduces the number of reflections contributing to the total signal. A further reduction of complexity of dynamical effects was expected from the 2BC.\footnote{Within the 2BC it was tried to work with a simple analytic model to describe the inelastic scattering using the transmitted beam and one Bragg scattered beam only. As it turned out, even in 2BC setup it is important to include other beams on the systematic row for quantitative EMCD description. And for higher thicknesses we need to abandon the systematic row approximation completely.}

Performing a measurement with incoming beam parallel to the zone axis leads to strong dynamical effects. A larger number of diffracted beams appears in the diffraction plane. The calculation becomes considerably more demanding, because many points of the reciprocal lattice are close to the Ewald sphere and often larger number of Bloch coefficient products need to be taken into account. Still, there is no conceptual difference in the treatment of the zone axis case or systematic row case.

The zone axis geometry has some advantages compared to 2BC or 3BC geometries. The beam tilt is zero, therefore the diffraction pattern may be of higher symmetry. Assuming a cubic structure of the sample and $(001)$ zone axis, the $lcc$ of the incoming beam can be either zero or lying along any of the symmetry axes ($\langle 100 \rangle$ or $\langle 110 \rangle$) and there will still be a mirror axis available in the diffraction plane allowing to separate out the imaginary part of MDFFs. E.g., we can arrange a 2BC-like geometry setting the $lcc$ to $\mathbf{G}/2$ [with $\mathbf{G}=(200)$ or $\mathbf{G}=(110)$ for bcc-Fe]. The higher symmetry of the setup allows to take advantage of available symmetry operations to improve the effective count rate and thus the SNR. When $lcc=(000)$ we have four symmetry operations (90$^\circ$ rotations along the transmitted beam) connecting possible detector positions to the symmetrically equivalent positions in the diffraction plane. This allows to symmetrize the signal and double the SNR. For the $(111)$ zone axis the diffraction pattern would acquire 3-fold rotational symmetry.

\begin{figure}[t]
  \includegraphics[width=8.5cm]{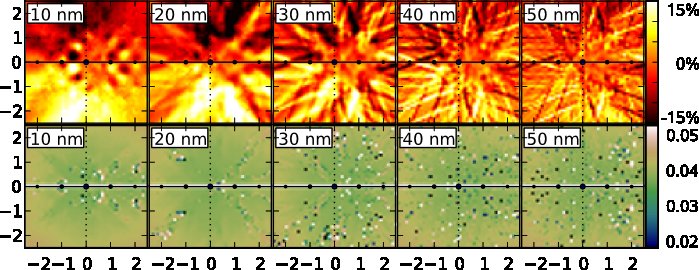}
  \caption{(color online) The same as fig.~\ref{fig:za3bc}, but with $lcc=(100)$, i.e., the 2BC-like geometry close to the ZA orientation. \label{fig:za2bc}}
\end{figure}

We performed two sets of zone axis simulations - an exact zone axis orientation (ZA), $lcc=(000)$ (see Fig.~\ref{fig:za3bc}), and a 2BC-like orientation (ZA-2BC) with $lcc=(100)$ (see Fig.~\ref{fig:za2bc}), respectively. Out of these, the ZA orientation is of higher symmetry - there are four mirror axes, on which the dichroic signal vanishes. All four mirror axes allow clean extraction of the magnetic part of the signal (no asymmetries), therefore it does not matter which one is chosen. Because the Thales circle positions for both $\mathbf{G}=(200)$ and $(110)$ are lying on the mirror axes, obviously these are not suitable for detector placement.\footnote{This information may also be important for experiments with small tilt angles, because then the experimental setup is rather close to ZA orientation and thus only a weak magnetic signal is expected. Moreover it would be heavily influenced by dynamical effects.} In the ZA-2BC geometry there is only one mirror plane left and the Thales circle positions may be, in principle, used. Though they are again not optimal, as can be seen from Fig.~\ref{fig:za2bc}.

In both cases the evaluated \ls{} ratio is smooth and constant throughout the diffraction plane, if one avoids areas with zero or low magnetic signal, where the error bar of \ls{} diverges. The signal intensity (and its magnetic content) naturally decreases with distance from the transmitted beam. Due to pronounced dynamical effects, the magnetic signal forms a complex pattern with many local minima and maxima, changing sign throughout the diffraction plane. This is a potential disadvantage of the ZA geometry, because the magnetic signal is low also in the surrounding of contour lines with zero magnetic signal. (This also limits the size of detector aperture.) By measuring the reciprocal maps it is possible to identify places with strong magnetic signal and use these areas for \ls{} evaluation. 

In ZA-2BC geometry the magnetic signal has less sign changes in the diffraction plane, therefore it might seem to be more suitable for experiments. Though, it would be desirable to study the effect of small misorientation, particularly the influence of nonzero component of $lcc$ perpendicular to $\mathbf{G}$. In ZA-2BC there is no additional mirror axis available, which would help to correct the map of magnetic signal by the double difference procedure.

\begin{figure}[t]
  \includegraphics[width=8.5cm]{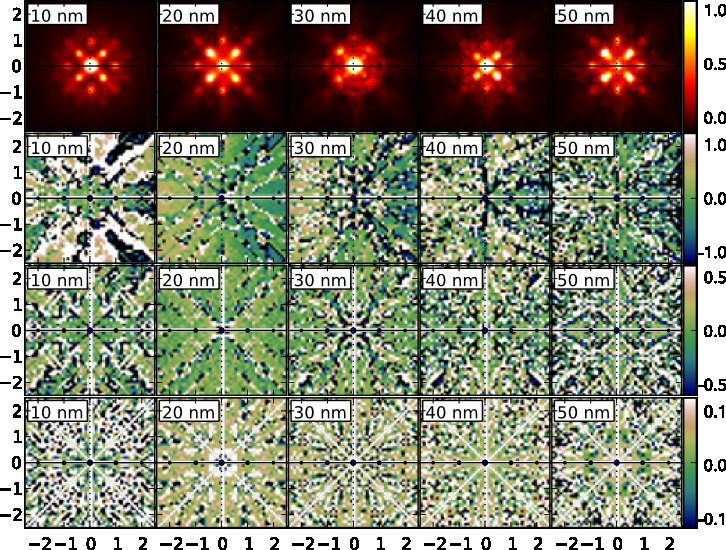}
  \caption{(color online) Slightly misoriented zone axis setup with $lcc=(0.07,0.04,0)$. Top row: diffraction patterns at $L_3$ edge (arbitrary units). Note the slight differences in intensities of $\{110\}$ Bragg spots, best visible at 30nm. Next rows show the apparent \ls{} ratio constructed using: horizontal mirror axis ($2^\mathrm{nd}$ row), double difference ($3^\mathrm{rd}$ row) and triple difference, i.e., using all available mirror axes ($4^\mathrm{th}$ row). Note, how the range of the color bar can be reduced, as we apply more symmetrization to the data.\label{fig:zat3bc}}
\end{figure}

In ZA orientation there are four mirror axes, which might be used to amplify the magnetic signal in difference maps in a way similar to the double difference procedure for 3BC. We tested it on a slightly tilted ZA orientation. The $lcc$ was set to $(0.07,0.04,0)$ - position, which is outside of any mirror axes. This introduces a slight asymmetry into the diffraction pattern, see Fig.~\ref{fig:zat3bc}. Asymmetry seems to be rather weak, it is best seen at thickness 30nm as an intensity variation among individual $\{110\}$ Bragg spots. Despite the fact that it is barely visible in the diffraction pattern, its influence on apparent \ls{} is very strong. See the map of apparent \ls{} ratio constructed from `up-down' difference maps in Fig.~\ref{fig:zat3bc}. It is obvious that this map would not allow to obtain reasonable quantitative magnetic information. A similar situation is obtained also for `left-right' based maps (not shown). The double difference map using both horizontal and vertical mirror axes considerably improves the map. Further improvement is achieved by exploiting also the diagonal mirror axes - a `triple difference'. Nevertheless, there are still quite strong deviations from the correct \ls{} value. 
As a whole, these simulations indicate that zone axis measurements of the \ls{} ratio are very sensitive to accurate orientation of the sample. Though, when the orientation can be set with sub-0.1~mrad accuracy, ZA measurements will provide highly symmetric maps and allow an accurate \ls{} extraction.

\section{Conclusions}

We explored the role of symmetry in EMCD experimental geometries, where the sample itself serves as a beamsplitter and the dichroic spectra are obtained as a difference from two detector positions. Various sources of asymmetries and their influences on the extracted \ls{} ratio are analysed. We have provided a survey of EMCD experimental geometries, stating their advantages and disadvantages. We have demonstrated, that the Thales circle positions are, in general, not optimal for detector placement. Optimal detector positions are sample-, thickness- and orientation-dependent.

Our analysis has shown, that the two-beam case is, strictly speaking, not suitable for the sum-rules application due to its inherent asymmetry. Despite this asymmetry, the dichroic signal can be rather strong and an accurate \ls{} evaluation can be performed within a reasonable range of detector positions. This geometry seems to be advantageous, when it is not possible to acquire energy-resolved images of the diffraction pattern and only spectra at particular detector positions are measured. We have shown that exact setting of the two-beam case is not crucial and deviations from two-beam case towards three-beam case do not qualitatively influence the \ls{} evaluation.

We have analysed the three-beam case, which enables two symmetric detector positions, provided there is a mirror symmetry plane perpendicular to the systematic row index. Possible deviations from perfect three-beam-case geometry can be corrected by constructing a double-difference map. This geometry might become the method of choice, when energy-resolved diffraction patterns are available. Information from the whole diffraction plane can be utilized\cite{klaus,lsfollow}.

Finally, as it turns out, measurement in the zone axis orientation is also possible. It brings an advantage of higher symmetry of the diffraction pattern, which may be used to improve the signal to noise ratio. On the other hand, pronounced dynamical diffraction effects cause a high sensitivity to even subtle deviations (few tenths of miliradians) from the exact zone axis orientation ($lcc=0$). These errors can be corrected to some extent by double or triple difference procedures, nevertheless in zone axis orientation this approach is not as efficient as in the three-beam case.

\begin{acknowledgments}
Stimulating discussions with Hans Lidbaum, Stefano Rubino and Klaus Leifer are gratefully acknowledged. Thanks go to Peter Schattschneider, Pavel Nov\'{a}k, Peter M. Oppeneer and Olle Eriksson, who had influenced this work at various stages, and to anonymous referees for their valuable comments. Data manipulation and plots were performed using R: A language and environment for statistical computing\cite{Rko}.
\end{acknowledgments}

\appendix

\section{Summation over Bloch coefficients\label{sec:dd12}}

Evaluation of the double differential scattering cross-section involves calculation of the following sum [see, e.g., Ref.~\onlinecite{prbtheory}, Eq.~(24)]
\begin{eqnarray} \label{dscsfin}
  \frac{\partial^2 \sigma}{\partial\Omega \partial E} & = & \sum_{\mathbf{ghg}'\mathbf{h}'}
    \frac{1}{N_\mathbf{u}} \sum_\mathbf{u}
                                           \frac{S_\mathbf{u}(\mathbf{q},\mathbf{q'},E)}{q^2 q'^2} e^{i(\mathbf{q}-\mathbf{q'})\cdot\mathbf{u}}
                      \nonumber \\  & \times &
    \sum_{jlj'l'} Y_{\mathbf{ghg}'\mathbf{h}'}^{jlj'l'} T_{jlj'l'}(t)
\end{eqnarray}
where
\begin{eqnarray}
  Y_{\mathbf{ghg}'\mathbf{h}'}^{jlj'l'} & = &
   C_{\mathbf{0}}^{(j)\star} C_{\mathbf{g}}^{(j)}
   D_{\mathbf{0}}^{(l)} D_{\mathbf{h}}^{(l)\star}
                       \\  & \times &
   C_{\mathbf{0}}^{(j')} C_{\mathbf{g'}}^{(j')\star}
   D_{\mathbf{0}}^{(l')\star} D_{\mathbf{h'}}^{(l')} .
                       \nonumber
\end{eqnarray}
Here $C_\mathbf{g}^{(j)}$ and $D_\mathbf{h}^{(l)}$ are the Bloch coefficients for the incoming and outgoing Bloch fields, $\mathbf{g},\mathbf{h}$ are Bragg reflections and $j,l$ are indices of the Bloch waves. The quantity $\frac{S_\mathbf{u}(\mathbf{q},\mathbf{q'},E)}{q^2 q'^2}$ is the MDFF divided by squares of momentum transfer vectors $\mathbf{q},\mathbf{q'}$ (Coulomb potential factors). For the rest of the notation we refer the reader to our above-mentioned paper.

The main difficulty originates in the eight-fold summation over Bragg reflections and Bloch-wave indices (to be performed $N_t \times N_E$ times, where $N_t$ is number of sample thicknesses and $N_E$ is number of energy-loss steps considered). Since the \ls{} ratio depends sensitively on accuracy of simulations, a large number of terms is needed to achieve good convergence of results. In our previous simulations a subset of Bragg reflections were selected by means of $w_{\mathbf{g},\mathrm{max}}$ and using these we selcected a subset of Bloch waves with certain minimum norm over these reflections. That was performed for both incoming and outgoig Bloch fields and all cross-terms between these indices were taken into the summation. This way a vast number of negligible terms were summed consuming valuable computing time.

Here we try to pre-identify the important terms and evaluate only those, which are bigger than some minimum value cut-off, $P_\mathrm{min}$, in this paper set to $10^{-5}$ (it may be increased or reduced, depending on accuracy demands). Since the thickness-dependent function $T_{jlj'l'}(t)$ has for given thickness $t$ moreless constant magnitude and the same is true for the phase factor $e^{i(\mathbf{q}-\mathbf{q'})\cdot\mathbf{u}}$, their precise values need not to be considered in estimation of term magnitudes. For the MDFF, for the sake of relative magnitude estimation, we assume a dipole type asymptotics, i.e.,
\begin{equation}
  \frac{S(\mathbf{q},\mathbf{q'},E)}{S(\mathbf{q}_0,\mathbf{q}_0,E)} \approx \frac{A \mathbf{q}\cdot\mathbf{q'} + \mathbf{B}\cdot (\mathbf{q} \times \mathbf{q'})}{A \mathbf{q}_0\cdot\mathbf{q'}_0} \lesssim \frac{q q'}{q_0^2}
\end{equation}
where $A$ and $\mathbf{B}$ are system dependent model parameters and we estimate them to be of similar magnitude. For non-magnetic centrosymmetric samples $\mathbf{B}=0$. Here $\mathbf{q}_0 = \mathbf{k}_\mathrm{out} - \mathbf{k}_\mathrm{in}$ and $\mathbf{q}_\mathbf{gh} = \mathbf{q}_0 + \mathbf{h} - \mathbf{g}$.

First we solve large eigenvalue problems for incoming and outgoing Bloch fields with Bragg-reflection cut-off condition $w_\mathbf{g}<10^5$. That is fulfilled by 600-700 Bragg reflections (in bcc Fe) and the same number of Bloch waves. From the resulting eigenvectors (Bloch waves) we construct a list of $C_\mathbf{0}^{(j)\star} C_\mathbf{g}^{(j)}$ and $D_\mathbf{0}^{(l)\star} D_\mathbf{h}^{(l)}$ products bigger than $P_\mathrm{min}/Q$, where $Q$ is another convergence parameter explained below. Number of elements in these lists varies between few thousands for systematic row geometries up to 2-3 tens of thousands elements in zone axis simulations. These lists are sorted by magnitude using the \textsc{QuickSort} algorithm.

The next step is generation of a list of quadruple products of Bloch coefficients, $C_{\mathbf{0}}^{(j)\star} C_{\mathbf{g}}^{(j)} D_{\mathbf{0}}^{(l)} D_{\mathbf{h}}^{(l)\star} q_0/q_{\mathbf{gh}}$, bigger than $P_\mathrm{min}$. This list is also sorted by \textsc{QuickSort} algorithm, but now by $\mathbf{q}$-vectors as the primary criterion (i.e., grouping the terms with the same $\mathbf{q}$-vectors next to each other) and magnitude as the second. Number of elements of this list is typically of the same order of magnitude as any of the lists of Bloch coefficient double-products constructed in previous step. Here it also becomes clear, what is the purpose of the parameter $Q$: since the magnitude of the ratio $q_0/q_{\mathbf{gh}}$ can in principle exceed unit, it may happen that a quadruple product with this factor will exceed $P_\mathrm{min}$ even if some of double products, of which it consists, are below this treshold. This may happen for larger scattering angles or higher energy losses and we adopt a very safe value $Q=10$. It is not too costly, since the lists of double products are typically smaller than other lists constructed in next steps.

Finally, a list of octuple products of Bloch coefficients $Y_{\mathbf{ghg}'\mathbf{h}'}^{jlj'l'} \frac{q_0^2}{q_\mathbf{gh} q_\mathbf{g'h'}}$ is generated. Thanks to the way, how the list of quadruple products is sorted, we can directly construct and output a $\mathbf{q},\mathbf{q'}$-ordered list of octuple products $Y_{\mathbf{ghg}'\mathbf{h}'}^{jlj'l'}$, along with all $\gamma^{(j)}, \eta^{(j)}$ (see Ref.~\onlinecite{prbtheory}), into a binary file for later summation. Generation of this list is linear in the number of generated octuple products.

Here we note that the factor $\frac{q_0^2}{q_\mathbf{gh} q_\mathbf{g'h'}}$ is the dipole-type asymptotic estimation of the relative magnitude of the MDFF with respect to the DFF, including Coulomb factors $1/q^2q'^2$. (A more strict estimation would be possible, since dipole approximation underestimates the decay of the dipole-allowed transitions for large $\mathbf{q}$ vectors.)

The list of $\mathbf{q,q'}$ is written to a file for an external program, which calculates the MDFFs. Obviously, the final summation is linear in the length of the list of octuple products, and is very fast and more accurate compared to a summation over all cross-terms performed before, because with this algorithm we can afford to include much larger number of beams and Bloch waves. The computing time scales moreless proportionaly to the inverse of $P_\mathrm{min}$, which is an enormous improvement over $N_\mathbf{g}^4 \times N_j^4$ of the previous method (per thickness and energy; $N_\mathbf{g}$ is number of Bragg reflections and $N_j$ is number of Bloch waves included in summation).

\begin{figure}[thb]
  \includegraphics[width=8.5cm]{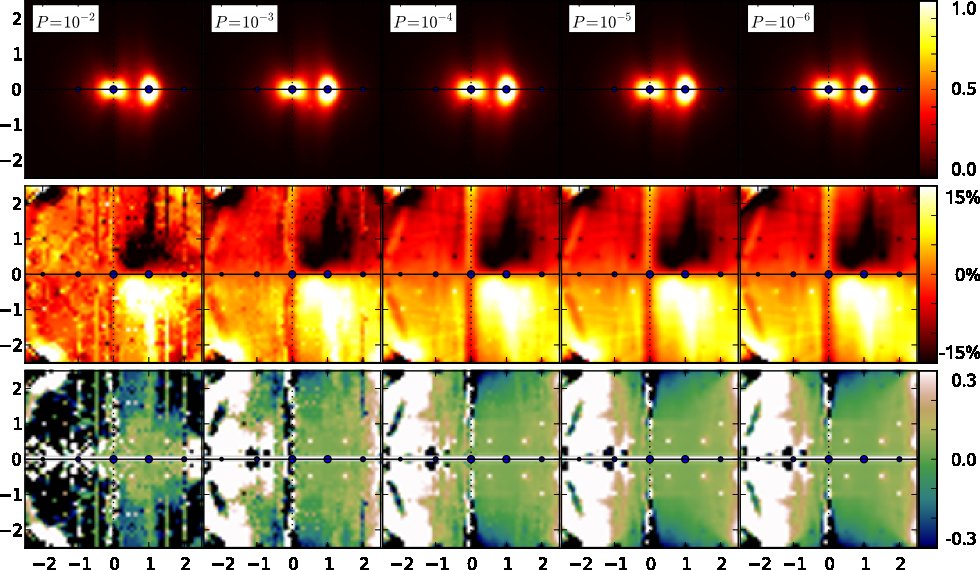}
  \caption{Diffraction patterns (top row), relative up-down difference maps (middle row) and apparent \ls{} ratio maps (bottom row) a function of convergence parameter $P_\mathrm{min}$. Calculations were performed for 20nm layer of bcc iron in two-beam geometry with $\mathbf{G}=(200)$, at 300keV.\label{fig:conv}}
\end{figure}

\begin{table}[thb]
  \begin{tabular}{lrrrrr}
    \hline \hline
 $P_\mathrm{min}$ & \multicolumn{1}{c}{$\langle N_2 \rangle$} & \multicolumn{1}{c}{$\langle N_4 \rangle$} & \multicolumn{1}{c}{$\langle N_8 \rangle$} & \multicolumn{1}{c}{$\langle N_\mathbf{qq'} \rangle$} & \multicolumn{1}{c}{$t$} \\
    \hline
      $10^{-2}$   &   230 &     72 &     330 &    12 &   2h 16min  \\
      $10^{-3}$   &  1355 &    750 &    4710 &   110 &   3h 18min  \\
      $10^{-4}$   &  5545 &   5865 &   56830 &   705 &  11h 13min  \\
      $10^{-5}$   & 23060 &  39760 &  570250 &  3600 &  83h 35min  \\
      $10^{-6}$   & 73591 & 259739 & 5398841 & 16860 & 628h 14min  \\
    \hline \hline
  \end{tabular}
  \caption{Average lengths of double $\langle N_2 \rangle$, quadruple $\langle N_4 \rangle$ and octuple $\langle N_8 \rangle$ product lists, average number of momentum transfer diads per energy step $\langle N_\mathbf{qq'} \rangle$ and computing times for maps in Fig.~\ref{fig:conv} as a function of convergence parameter $P_\mathrm{min}$. Times refer to a single Intel Pentium 4 Xeon processor at 2.5GHz.\label{tab:conv}}
\end{table}

We summarize the properties of the described algorithm in the Fig.~\ref{fig:conv} and Table~\ref{tab:conv} for five different values of convergence parameter $P_\mathrm{min}$. The calculation is performed on a grid of $51 \times 51$ outgoing beam directions for $N_E=10$ and $N_t=100$ in the two-beam orientation. Note that the $N_t$ only marginally influences the total computing time, since it only scales the final summation, which is typically the fastest part of the calculation. The $N_E$ influences linearly the MDFF calculation and final summation time. Generation of octuple products and $\mathbf{q,q'}$ lists is $N_t$ independent and only marginally $N_E$ dependent (it only influences the output of $\mathbf{q,q'}$ list).

Note that the diffaction pattern appears to be reasonably converged already for $P_\mathrm{min}=10^{-2}$. An attentive reader might spot slightly fuzziness, but nevertheless, the difference between all five diffraction patterns are visually negligible. The relative difference map requires better convergence, at least $P_\mathrm{min}=10^{-3}$, or better $P_\mathrm{min}=10^{-4}$. Note the numerical noise at $P_\mathrm{min}=10^{-3}$, particularly along the vertical line going through the transmitted beam. Although a slight fuzziness remains at $P_\mathrm{min}=10^{-3}$, but the accuracy is already satisfactory. The most sensitive quantity presented here is the apparent \ls{} ratio. Since the orbital momentum in iron is small, the nominator in the sum rule expression is a difference of two (typically small) differences of spectra, integrated over $L_2$ or $L_3$ edge, respectivelly.
It requires high accuracy to obtain well converged maps. Calculation with $P_\mathrm{min}=10^{-4}$ seems to produce reasonably converged results. The only visible difference at $P_\mathrm{min}=10^{-5}$, when compared to $P_\mathrm{min}=10^{-4}$, appears around $(-1,\pm 1)G_{(200)}$ positions. The results of calculation with $P_\mathrm{min}=10^{-6}$ are visually indistinguishable from $P_\mathrm{min}=10^{-5}$. So we can conclude that for purposes of this paper the convergence criterion $P_\mathrm{min}=10^{-5}$ provides highly converged data - particularly when we consider a region of sizable difference signal, where an appreciable the magnetic signal to noise ratio can be expected.

The Table~\ref{tab:conv} summarizes somewhat more technical information about the computational complexity. We show the average lengths of lists of Bloch-coefficient products, which scale moreless inverse-proportionally to the convergence parameter. The computing time apparently grows slowlier, but one should take into account a constant everhead in these calculations connected with two diagonalizations of complex matrices of dimension ca.~630 and the start-up and initialization costs of MDFF calculation, which are both repeated $51 \times 51$ times. These overheads are estimated to be roughly around 2 hours. After subtraction of that a moreless linear scaling of the computation time with inverse $P_\mathrm{min}$ is observed. We note that the computing time is typically by factor 10-15 longer for zone axis calculations.

The whole algorithm is trivial to parallelize, since each pixel in the diffraction plane is calculated independently. Therefore practically a linear scaling with number of processors is achieved. For applications, where only the diffraction pattern is needed, the method delivers accurate results at affordable computing time.

\section{A note on optimal detector positions\label{sec:dichmaps}}

An important corollary of the sum-rules is, that the imaginary part of energy integrated MDFF does contain the magnetic information regardless of the relative orientation of $\mathbf{q}$ and $\mathbf{q'}$ momentum transfer vectors. The only exceptions are parallel or antiparallel orientations, respectively, when the imaginary part is zero (within dipole approximation). This implies, that the detector placement is by no means limited to the Thales circle positions (Fig.~\ref{fig:assym}). In other words, the Thales circle positions have no special importance\footnote{This relates to experiments described here, where the crystal itself is used as a beamsplitter.}. We have presented a set of simulations varying the detector position to scan a part of the diffraction plane, which lead to difference maps related to magnetic signal. These maps provided several findings: i) the maximum of the dichroic signal is, as anticipated, at places other than at the Thales circle positions; ii) there can be several local maxima and minima in the diffraction plane; iii) the pattern of the dichroic signal sensitively depends on the sample thickness and orientation.

This means, that the question of the optimal detector positions actually does not have a unique general answer. There are orientations and thicknesses, when the optimal detector positions are quite different from Thales circle positions. All this is also sample dependent. Therefore in actual experiments it would be advantageous to have access to such maps of the dichroic signal. Such experiments actually have been already done.\cite{toulouse,klaus,lsfollow}

Another application of the above-mentioned corollary of sum-rules is the possibility to integrate the dichroic signal over some part of the diffraction plane. Because the prefactors in spin and orbital sum-rules are the same, the signal from a range of detector positions can be collected. This can be done either by numerically integrating within the map of dichroic signal\cite{snr} or by having a suitable collection aperture. This observation is related to the question of optimizing the SNR - selection of the optimal size, shape and placement of the spectrometer entrance aperture.\cite{snr} [Alternative approach of improving SNR of a measured physical quantity - \ls{} ratio - was developed in Ref.~\onlinecite{klaus}, where a statistical analysis was performed within \ls{} maps.]

\section{Energy integration range\label{sec:eint}}

\begin{figure}
  \includegraphics[angle=270,width=8.5cm]{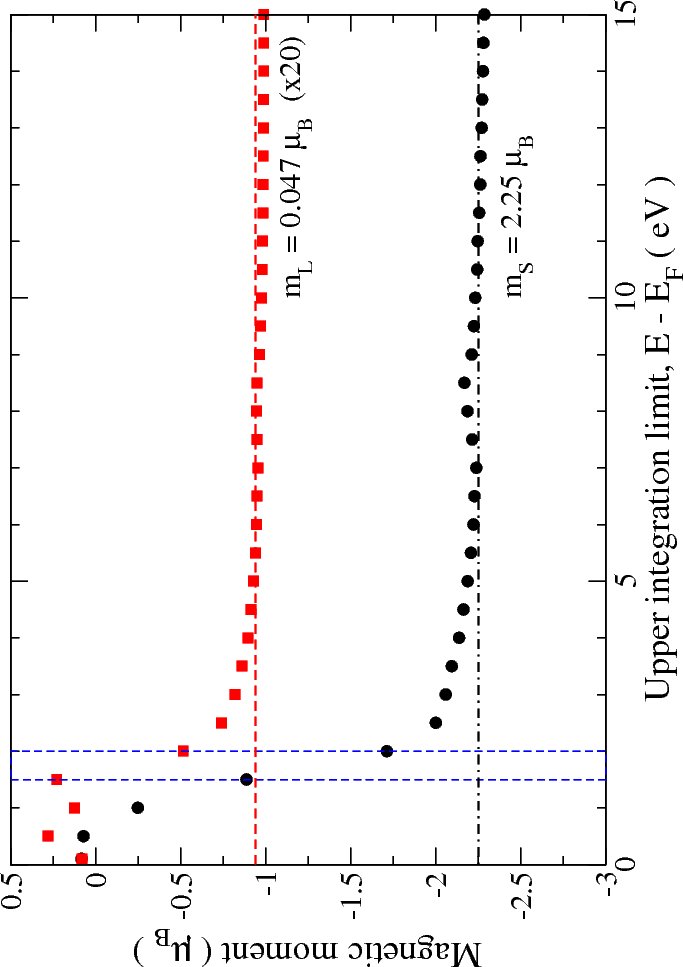}
  \caption{(color online) Mean value of the spin and orbital operators for unoccupied states as a function of the upper integration limit $E$. The energy range inside the blue dashed rectangle was considered in calculations. Ground state values are marked by horizontal lines.\label{fig:fels}}
\end{figure}

Extensive calculations presented in this work would not be feasible without approximations, which were presented in the description of the computational methods. These approximations were carefully chosen to provide correct unbiased results. There is one exception though, which is the energy integration range in \ls{} evaluation. This should be in principle infinite, $(E_F,\infty)$. Since the MDFF evaluation is the most time-consuming step of calculations, the energy integration range was reduced to a small interval $(E_F+1.5\mathrm{eV},E_F+2.0\mathrm{eV})$ with 11 equally spaced energy values. This interval was chosen to encompass the main peak in the unbroadened spectrum and applies for both $L_2$ and $L_3$ edges. Consequences of this approximation are briefly discussed in this appendix.

In Fig.\ref{fig:fels} we present the expectation values of the spin and orbital momentum operators evaluated as a function of the energy integration range, $(E_F,E_F+E)$, within the $d$-states subspace. For infinite $E$ these values would converge to ground state spin and orbital moments of $d$-states, which are represented by horizontal lines. The evolution of both spin and orbital momentum as a function of energy is different, therefore also the `apparent' \ls{} ratio will depend on the energy range. Particularly, for the range $(E_F+1.5\mathrm{eV},E_F+2.0\mathrm{eV})$ the \ls{} ratio is $0.045$ -- that is roughly twice the ground state value.

It is important to realize that this fact does not influence any of the previous qualitative conclusions, which are system-independent. Indeed, the maps of dichroic signal as a function of energy show identical trends with varying amplitudes.

\section{Double difference correction\label{sec:dd}}

\begin{figure}
  \includegraphics[width=8.5cm]{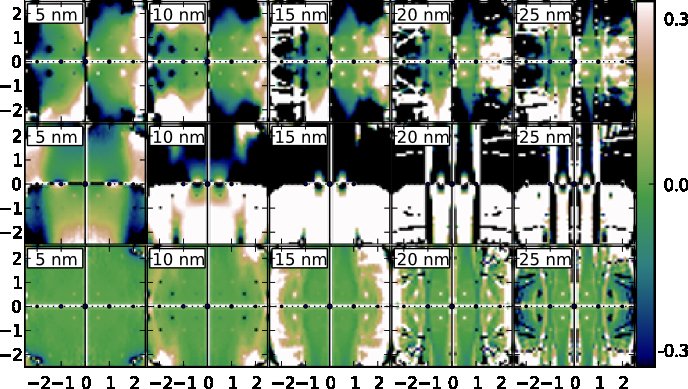}
  \caption{(color online) Map of the $2\Delta_\mathrm{Re}/\Delta_\mathrm{Im}$ for a slightly tilted 3BC, $lcc=(0.05,0,0)$. Top row: `up-down' difference (horizontal mirror axis), middle row: `left-right' difference (vertical mirror axis), bottom row: double difference (both mirror axes).\label{fig:re23bc} }
\end{figure}

The `left-right' difference signal extracted from a perfect 3BC geometry was shown to be of purely magnetic origin. In practice, though, there is always some deviation from perfect 3BC towards $\mathbf{G}$ or $-\mathbf{G}$. [The beam tilt leading to a systematic row setup is of much larger magnitude ($\sim 0.1$ rad) than tilts discussed here (sub-miliradian). Therefore deviations along $k_y$ do not play role--they are simply absorbed into the beam tilt--and one only has to deal with deviations along $k_x$ (the $x$ axis is along the systematic row).] It was demonstrated in Fig.~\ref{fig:lsdd} that double difference correction significantly extends the area, where an accurate $m_L/m_S$ ratio extraction is possible. Here we will briefly discuss, why this correction works.

By slightly tilting the beam towards positive $k_x$, the intensity throughout the diffraction pattern increases in the right half-plane at all energy levels. Particularly, this happens for the nonmagnetic component of the intensity. We assume that the increase is similar in upper and lower half-planes, i.e., $\Delta I(k_x,k_y) \approx \Delta I(k_x,-k_y)$, while $\Delta I(-k_x,k_y)$ is of different sign. Then the left-right difference will contain an error proportional to $\Delta I(k_x,k_y)-\Delta I(-k_x,k_y)$, which increases the magnitude of magnetic signal on one edge and decreases it on the other edge (assuming that they are of different sign). This biases the observed \ls{} value. Additionally, this error influences the upper and lower half-plane in opposite ways and thus the extracted \ls{} ratio differs in upper and lower half-planes. 

If we perform a `left-right' difference, the nonmagnetic components from the exact 3BC orientation cancel out and only the ``error'' superimposed on magnetic signal remains. Subsequent `up-down' difference enhances the magnetic signal and, simultaneously, reduces the error thanks to its approximate up/down symmetry mentioned above: $\Delta I(k_x,k_y) \approx \Delta I(k_x,-k_y)$. Remaining error, $[\Delta I(k_x,k_y)-\Delta I(k_x,-k_y)]-[\Delta I(-k_x,k_y)-\Delta I(-k_x,-k_y)]$, is considerably smaller than asymmetry in pure `up-down' or `left-right' differences, respectively.

To verify these assumptions quantitatively we calculated separately the magnetic and nonmagnetic contribution to the scattering cross-section for the slightly misoriented 3BC with $lcc=(0.05,0,0)$. This amounts to two separate calculations considering either only real parts (non-magnetic component) or imaginary parts (magnetic contribution) of MDFFs, respectively.

The difference maps of the non-magnetic part of the signal ($\Delta_\mathrm{Re}$; not shown) are directly related to the error introduced by asymmetries. Very instructive is to show a ratio $\Delta_\mathrm{Re}/\Delta_\mathrm{Im}$, i.e., the difference of non-magnetic signal divided by the corresponding difference of the magnetic signal. Such maps describe the relative enhancement of the apparent magnetic signal due to non-magnetic contribution. Maps of $\Delta_\mathrm{Re}/\Delta_\mathrm{Im}$ evaluated at the $L_3$ edge are shown in Fig.~\ref{fig:re23bc} for `left-right', `up-down' and double difference, respectively. Maps at $L_2$ edge (not shown) have the same trends, but with opposite sign and reduced magnitude. Note that values in the double difference map are much lower than in the other two maps, demonstrating the effectiveness of the the double difference procedure.

Note that these maps are strikingly similar to Fig.~\ref{fig:lsdd}. It should not be surprising though, since they quantify the error caused by asymmetry. The error introduced into \ls{} evaluation due to asymmetry can be expressed more precisely in terms of the following equations. Let's assume that the errors $\Delta_\mathrm{Re}$ at $L_3$ are twice as large compared to errors at $L_2$ edges. This is approximately the case for late $3d$ metals, where the branching ratio is close to 2. Then the apparent \ls{} ratio is
\begin{eqnarray*}
  \lefteqn{\frac{2}{3} \frac{(\Delta_{L_3}+2\Delta_\mathrm{Re})+(\Delta_{L_2}+\Delta_\mathrm{Re})}{(\Delta_{L_3}+2\Delta_\mathrm{Re})-2(\Delta_{L_2}+\Delta_\mathrm{Re})} = } \\
 & = & \frac{2}{3} \frac{\Delta_{L_3}+\Delta_{L_2}+3\Delta_\mathrm{Re}}{\Delta_{L_3}-2\Delta_{L_2}} \\
 & = & \frac{m_L}{m_S} + \frac{2\Delta_\mathrm{Re}}{\Delta_{L_3}-2\Delta_{L_2}} \approx \frac{m_L}{m_S} + \frac{2\Delta_\mathrm{Re}}{3\Delta_\mathrm{Im}}
\end{eqnarray*}
where $\Delta_{L_{2,3}}$ stands for $\Delta_\mathrm{Im}$ at particular edge. In the last step we assumed that $\Delta_{L_3} \approx -\Delta_{L_2} \equiv \Delta_\mathrm{Im}$. This equation expresses the quantitative influence of the error caused by asymmetry in terms of the ratio $\Delta_\mathrm{Re}/\Delta_\mathrm{Im}$, and explains the similarity of Fig.~\ref{fig:re23bc} and Fig.~\ref{fig:lsdd}. We note that the error is \emph{additive}.

\end{document}